# Photonic Microwave and RF Channelizers using Kerr Micro-combs

Mengxi Tan[1], Xingyuan Xu[1,2], and David J. Moss[1] [†]

[1] Optical Sciences Centre, Swinburne University of Technology, Hawthorn, VIC 3122, Australia.
[2] Electro-Photonics Laboratory, Dept. of Electrical and Computer Systems Engineering, Monash University, VIC3800, Australia.
[†] Corresponding e-mail: dmoss@swin.edu.au

*Abstract*— **We review recent work on broadband RF channelizers based on integrated optical frequency Kerr micro-combs combined with passive micro-ring resonator filters, with microcombs having channel spacings of 200GHz and 49GHz. This approach to realizing RF channelizers offers reduced complexity, size, and potential cost for a wide range of applications to microwave signal detection.**

## I. INTRODUCTION

Nonlinear optics as a means to achieve ultrafast all-optical signal processing has been extremely powerful, and has focused on photonic integrated circuit platforms based on highly nonlinear materials such as silicon [1-3]. The range of signal processing operations that can be performed with nonlinear optical techniques is very large, and includes optical logic [4], optical temporal demultiplexing at ultra-high bit rates from 160Gigabits/s [5] to well over a terabit per second [6], optical performance monitoring based on slow light [7,8], signal regeneration [9,10], and many other functions [11-16]. Complementary metal oxide semiconductor (CMOS) compatible platforms are extremely important since they can exploit the extensive global infrastructure established to fabricate computer chips. Since all CMOS compatible material platforms are centrosymmetric, the 2$^{rd}$ order nonlinear response is zero and so nonlinear devices in these platforms have been based on the 3$^{rd}$ order nonlinear susceptibility that accounts for processes including optical third harmonic generation [11,17-21] as well as the instantaneous intensity dependent refractive index termed the Kerr nonlinearity ($n_2$) [1,2]. The efficiency of all-optical devices based on the Kerr nonlinearity depends on the waveguide nonlinear parameter ($\gamma$), and although silicon waveguides and nanowires have achieved extremely high nonlinear parameters ($\gamma$), they also display quite high nonlinear optical loss arising from indirect two-photon absorption (TPA) in the telecommunications wavelength region (near 1550nm), since the indirect bandgap of silicon is only 1.17eV which is much less than twice the photon energy at 1550nm. Moreover, worse than the intrinsic TPA is the effect of the resulting generated free carriers [2] since the carrier recombination time in silicon can be very long (microseconds), enabling significant carrier densities (and hence loss) to build up. Even if the effect of these carriers can be minimized through the use of p-i-n junctions to sweep out the carriers and greatly reduce the effective carrier lifetime, the intrinsic material nonlinear figure of merit (FOM = $n_2 / (\beta\,\lambda)$, where $\beta$ is the TPA and $\lambda$ the wavelength) of silicon is only 0.3 in the telecom band. This is too low to achieve high all-optical performance and is a fundamental property of silicon's bandstructure and therefore cannot be improved. While TPA can be turned around and actually used as a positive tool for some nonlinear functions [22-24], for most processes silicon's low FOM in the telecom band poses a significant limitation. This has been the motivation for research into alternative nonlinear optical platforms, such as chalcogenide glass [25-34], for example. However, although many of these platforms offer significant advantages, most of them are not CMOS compatible, which ultimately is a fundamental consideration to achieve widespread and efficient manufacturing.

New CMOS compatible platforms for nonlinear optics were introduced in 2007/8 that exhibit negligible TPA in the 1550nm wavelength regime. These include silicon nitride [35, 36] and high-index doped silica glass, trade-named Hydex [37-47]. In addition to having negligible nonlinear absorption even up to many

Gigawatts of power, these platforms have a moderate Kerr nonlinearity, resulting in an extremely high nonlinear FOM with a nonlinear parameter (γ) that is high enough to realize significant parametric gain. Following the first report of a micro-resonator based optical frequency comb source driven by the Kerr nonlinearity in 2007 [48], the first fully integrated optical parametric oscillators based on micro-ring resonators (MRRs) were reported in 2010 [36, 37] in these new CMOS compatible platforms of SiN and Hydex. Since 2010 the field of integrated micro-combs, or Kerr combs has become one of the largest and most successful fields in optics and photonics [47]. They are a new and powerful tool to accomplish many new functions on an integrated chip, due to their very high coherence, while also offering flexible control of their wavelength spacing. Optical micro-combs are produced via optical parametric oscillation driven by modulational instability, or parametric, gain in integrated ring resonators. This offers huge advantages over more conventional approaches including discrete multiple laser wavelength sources. Many breakthroughs have been reported with Kerr micro-combs, from innovative mode-locked lasers [49-52] to quantum optical photonic chips [53-61], ultrahigh bandwidth optical fiber data transmission [62-64], optical neural networks [65-67], integrated optical frequency synthesizers [68] and more, and have been extensively reviewed [47, 69 - 76]. The success of these new CMOS platforms has motivated the search for even higher performing CMOS compatible platforms including as amorphous silicon [77] and silicon rich silicon nitride [78], searching for the ideal combination of low linear and nonlinear loss together with a high nonlinearity.

A significant application for microcombs has been for signal processing functions for telecommunications and RF/microwave systems, including RF photonic applications to signal generation and processing for radar systems [79-128]. RF photonics is attractive because of its ultra-high bandwidth as well as low transmission loss and strong immunity to electromagnetic interference. There are different RF photonic approaches including methods that map the optical filter response onto the RF domain including devices based on stimulated Brillouin scattering [88-95], that have achieved high RF resolution — down to 32 MHz, a stopband discrimination > 55 dB. A different but equally key approach to reconfigurable RF transfer functions for signal processing has been transversal filter methods [96-100]. These operate by generating progressively delayed and weighted replicas of an RF signal multicast onto many optical carriers, which are subsequently summed via photo-detection. Transverse filters can achieve a wide range of RF functions solely by varying the tap weights, and so this approach is very attractive for advanced dynamically adaptive RF filters. Discrete diode laser arrays [101] and fibre and integrated Bragg grating arrays and sampled gratings [103] have been used to generate the required taps. However, while offering advantages, these methods have increased complexity and footprint, limiting performance due to the limited number of wavelengths. Alternative methods such as electro-optic (EO) or acousto-optic (AO) combs [102,104,105], can help overcome this, but they require many high bandwidth modulators and high-frequency RF sources.

Kerr micro-combs [36, 37] have been very successful in their applications to RF systems, providing advantages over other methods of supplying multiple wavelengths for RF photonic systems. They have achieved extremely high bandwidth data communications as well as a wide range of microwave signal processing devices [107-128]. Their comb spacings can be much wider than electro-optic combs, and in many ways EO and micro combs are complementary. EO combs excel at finer spacings from 10's of megahertz to 10 - 20 GHz, while integrated micro-combs typically have much wider spacings from 10's of GHz to 100's of GHz and even THz. Larger comb spacings have much wider Nyquist zones for large RF bandwidth operation, whereas smaller spacings provide many more wavelengths or RF "taps", although at the expense of a smaller Nyquist zones. Micro-combs provide more wavelengths while still having a large FSR, all in a small footprint. For RF transversal filters the number of wavelengths determines the number of channels for RF true time delays and RF filters [85, 121]. Systems such as beamforming devices [112] can also be improved in quality factor and angular resolution. Other approaches to filtering include RF bandwidth scaling [125] that yields a particular bandwidth for each wavelength channel, with the total bandwidth (maximum RF signal bandwidth) will depend on the number of wavelengths which is dramatically increased with micro-combs.

Recently [121], we reviewed transversal filtering and bandwidth scaling methods based on Kerr micro-combs, as applied to RF and microwave spectral filters followed by a review of temporal based signal processing [129]. In this paper, we review recent progress in microwave and RF photonic based channelizers based on Kerr micro-combs. We cover devices achieved with both widely spaced micro-combs with an FSR of 200GHz [113] as well as results obtained with record low FSR micro-combs with a spacing of 49 GHz, operating via soliton crystals [122]. We highlight their potential and future possibilities, contrasting the different methods and use of the differently spaced micro-combs. While 200GHz Kerr micro-combs have been successful for RF channelizers, achieving high levels of performance, high versatility and dynamic reconfigurability, their large comb spacing limited the number of channels as well as the RF bandwidth that can be achieved without the use of thermal tuning. This is an important consideration because RF channelizers need many components such as optical amplifiers and spectral shapers that are only available at telecom wavelengths (1530-1620nm). The limitation in the number of channels has restricted the overall RF bandwidth, frequency resolution, and dynamic reconfigurability of micro-comb based RF channelizers. To overcome this, we focused on RF channelizers based on record low FSR combs, with a spacing of 49GHz, in order to achieve over 90 channels in the C-band [122]. This represents the highest for any micro-comb based RF channelizer. Our results confirm the feasibility of achieving high performance reconfigurable transversal RF filters for signal processing with reduced footprint, complexity, and cost. We first review the microcombs used to realize the RF channelizers, followed by the results for 200GHz microcomb based channelizers and finally by the 49GHz soliton crystal based systems.

## II. INTEGRATED KERR MICRO-COMBS

The formation of micro-combs is a complex process that arises from a combination of a high nonlinear parameter together with a low nonlinear loss as well as a low linear loss, and finally with careful engineering of the dispersion. Micro-combs have been realized in a variety of material platforms [47] including magnesium fluoride, silicon nitride, silica, and doped silica glass [47, 70, 85]. In 2008, [39] efficient FWM operating at low (milliwatt) CW power levels was reported in a MRR with a 575GHz FSR spacing, in a ring resonator with a low Q-factor of around 60,000. This was the first report of any form of nonlinear optics at milliwatt CW power levels in a silica glass-based platform. In 2010 this was followed by the first report of an integrated Kerr micro-comb [37, 38]. Another key breakthrough came in 2017 [116, 117] with reports of integrated Kerr micro-combs with record small FSR's under 50GHz. This significantly increased the available number of channels, or wavelengths, to more than 80 in the telecommunications wavelength C band from 1530-1565nm. In addition to their low FSR spacings, these Kerr micro-combs operated in a different mode to DKS states [66-73], in a process that has been called soliton crystals [130, 131]. Many other innovative states have been reported in micro-combs, including microcombs with extremely low threshold powers [132], dark solitons [133], laser-cavity solitons [134] and others [135-140].

The devices that were used in the work reviewed here were fabricated in Hydex glass [37, 38], a CMOS compatible platform categorized as high index doped silica. Micro-ring resonators with Q factors ranging from 60,000 to more than 1.5 million have been achieved, with radii ranging from 592 μm down to 48 μm, corresponding to FSRs from 49 - 575 GHz. The RF signal processors used as the basis for the work reviewed here were fabricated in MRRs with FSR spacings of 200GHz and 49GHz. Films of Hydex glass ($n$ = ~1.7 at 1550 nm) were deposited via PECVD and photolithographically patterned with deep ultraviolet stepper mask aligners. Waveguides with very low surface roughness were created by reactive ion etching, after which an upper cladding of silica glass ($n$ = ~1.44 at 1550 nm) was grown. We use both lateral and vertical bus-ring coupling schemes, with a gap typically of about 200nm. Vertical coupling can be controlled by film growth more accurately than by lithography. Hydex glass has many advantages including a very low linear loss (~ 0.06 dB·cm$^{-1}$), a reasonably large nonlinear parameter (0.233 W$^{-1}$·m$^{-1}$), and probably most importantly, negligible TPA even up to many gigawatts cm$^{-2}$ of power. We achieved high Q factor resonators of ~1.5 million for both 49GHz and 200GHz spaced MRRs. After packaging with fiber pigtails, the coupling loss can be as low as 0.5 dB per facet with the aid of on-chip mode converters.

For the 200GHz devices, combs were generated by amplifying the CW pump power to more than 1W (> +30 dBm). Next the wavelength was tuned from blue to red relative to the resonance wavelength, targeting a TE resonance near ~1550 nm. As the offset between the cold resonance and pump wavelength became sufficiently small the intracavity pump power reached a threshold when modulation instability gain yielded oscillation [47], first generating primary combs with its spacing determined by the MI gain peak, which is a function of dispersion as well as the intra-cavity pump power. As the detuning changed further, microcombs with a spacing equal to the FSR appeared. While these were not soliton states, nonetheless we found that they were more than adequate for our applications to microwave signal processing. We found that it was not necessary to strictly operate in soliton states such as the dissipative Kerr solitons (DKS) [66]. This is an important issue since, while much progress has now been made around DKS solitons [140], they nonetheless intrinsically require quite complicated pump tuning dynamics, generally employing both simultaneous amplitude (power "kicking") and wavelength sweeping, including in reverse directions, to be able to "kickstart" the solitons from out of the chaos. Our early work on micro-combs was based on these 200GHz FSR combs that operated in this partially coherent state. As mentioned, while not rigorous soliton states, they were still low noise enough and managed to avoid the chaotic regime [47] - we found them to be more than adequate for our RF work. They were used to successfully demonstrated many different RF signal processing functions [85, 107, 109 -111].

More recently, we have used soliton crystal microcombs that featured not only this new mode of operation, but with a record low FSR spacing of 49GHz [130, 131]. Soliton crystals arise from mode crossings and are easier and more reliable to generate than other solitons, including DKSs, and even easier than the partially coherent states of the 200GHz resonators. They can be generated even with simple manual control of the pump wavelength and power, without any complex pump dynamics. The underlying physics for this is based on the fact that the internal optical energy in the cavity of the soliton crystals is very close to that of the chaotic state. Hence, when soliton crystals are generated from chaotic states, there is only a small net change in the intracavity optical energy. This means that there is virtually no induced shift in the resonant wavelength – either thermally or via the Kerr nonlinear effect. It is this shift in resonant wavelength and internal energy that makes DKS states so difficult to generate since the resonance "avoids' the pump wavelength via the self-induced shifting. This same effect is also responsible for greatly increasing the efficiency of the soliton crystals, so that the energy in the comb lines are much higher relative to the pump power than the DKS states, particularly for single soliton DKS states. There is one drawback of soliton crystal states, however, and that is that their spectra are not flat – they have characteristic nonuniform "curtain" patterns. While this can sometimes mean that spectral flattening is required, this is not always the case. Indeed, this issue has not prevented soliton crystals from achieving many breakthroughs in high performance photonic RF microwave functions. Even though the resonator mode-crossing needs to be engineered, as well as anomalous dispersion, these issues have not posed a significant barrier and high fabrication yields have been achieved [64]. The specific micro-comb states reached was not important for RF applications, only that low RF noise and high coherence be achieved, and we found this to be easy to generate with simple pump wavelength tuning. Soliton crystals have yielded the lowest noise levels of any of the micro-comb states that we have worked with and so we have focused on them for our RF work, featuring a low phase-noise microwave oscillator [120].

## III. PHOTONIC RF AND MICROWAVE CHANNELIZERS

The ability to detect and analyze RF and microwave signals with very large bandwidths is critical for radar systems, electronic warfare, satellite communications and more [141-150]. RF and microwave channelizers, which effectively slice the RF spectrum into many frequency bands that are compatible with digital processing [151], are attractive for RF signal detection and analysis. Conventional RF channelizers usually use an array of RF filters and so are susceptible to the electronic bandwidth limitation. Photonic methods are promising for RF channelizers since they can achieve high bandwidths and also have very strong electromagnetic interference immunity.

Early reports of RF photonic channelizers used methods where RF signals were transmitted over single optical wavelengths and were then physically split to achieve spectral channelizing, usually with diffraction gratings [12], acousto-optic crystals [153], fiber Bragg gratings [154], or integrated photonic chips [155]. However, all of these methods require many narrow linewidth and spectrally dense, precisely centered, filters. Hence these systems tend to be limited in spectral resolution as well as in the number of RF channels. They also tend to have a large footprint.

Recent work on RF spectral slicing using multicasting of RF signals onto many optical wavelengths simultaneously has adopted a number of techniques including stimulated Brillouin scattering [156-158], parametric nonlinear optical processes in fiber [159, 160], spectrally incoherent sliced sources [161], discrete laser arrays [162], or electro-optic modulator generated frequency combs [151, 163, 164]. Other approaches that use wavelength scanning devices [165] or dispersive Fourier transformations [166] have also been reported, but these generally face limitations of one sort or another, including limited channel numbers and limited RF spectral resolution. They also tend to be fairly complex and expensive because of the requirement for many components such as external RF sources and mode-locked lasers etc.. Integrated Kerr micro-combs [36-38, 47, 167], particularly those fabricated in CMOS compatible platforms [37- 47, 49 - 61, 168, 169], have many advantages for broadband RF channelizers compared with discrete multi-wavelength sources, such as being able to provide a much higher number of channels or wavelengths [107-129, 170] with a greatly reduced footprint and lower complexity.

Here, we review recent work on broadband photonic RF channelizers realised through the use of integrated optical frequency Kerr microcombs, in combination with passive ring resonator filters. By using an on-chip Kerr comb source consisting of an active nonlinear micro-ring resonator (MRR) with an FSR spacing of either 200GHz or 49GHz, in combination with a passive on-chip MRR based filter with a free spectral range (FSR) of 49 GHz and Q factor of $1.55 \times 10^6$, we achieved RF channelizers with high performance. The use of a 49GHz comb resulted in a larger number of channels since the FSR of the comb source was nearly matched to the 49GHz passive MRR. For the 200GHz spaced comb, only every 4th resonance of the 49GHz passive MRR could be used. We verified the RF performance experimentally at frequencies up to 26 GHz, achieving a high spectral resolution of better than 1.04 GHz for both systems. In addition to high RF performance, this approach offers a reduced footprint, lower complexity, and potentially lower cost.

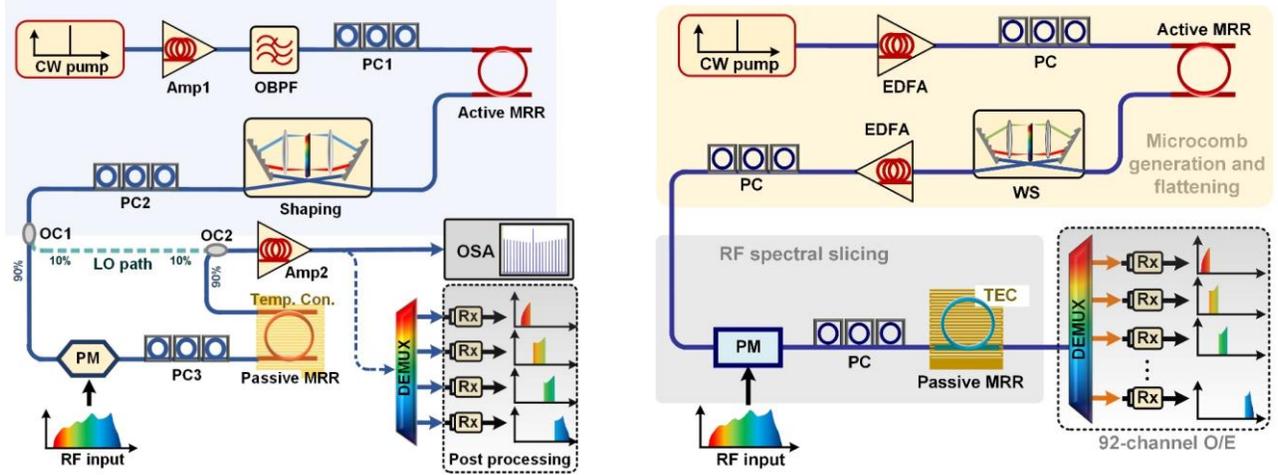

Fig. 1. Schematic diagram of the broadband RF channelizer based on an integrated optical comb source. Amp: erbium-doped fibre amplifier. OBPF: optical bandpass filter. PC: polarization controller. MRR: micro-ring resonator. OC: optical coupler. PM: phase modulator. Temp. Con.: temperature controller. DEMUX: de-multiplexer. Rx: Receiver. OSA: Optical spectrum analyzer. Right: Schematic diagram of the broadband RF channelizer based on a soliton crystal microcomb. EDFA: erbium-doped fibre amplifier. PC: polarization controller. MRR: micro-ring resonator. WS: Waveshaper. PM: phase modulator. TEC: temperature controller. DEMUX: de-multiplexer. Rx: Receiver.

## IV. PRINCIPLE OF OPERATION

Figure 1 shows a schematic of the RF channelizer operating by combining an active microcomb as a source (in our case combs with FSRs of either 200GHz or 49GHz) with a 49GHz passive MRR that acts as a narrow optical filter. We first discuss the system based on the 200GHz FSR microcombs, where Kerr combs were generated in the MRR by pumping by a tunable CW laser, amplified by an erbium-doped fibre amplifier. A tunable optical bandpass filter suppressed the amplified spontaneous emission noise from the amplifier, and a fibre polarization controller was used to adjust the polarization to optimize the coupled power into the waveguide. When the pump wavelength was tuned near a resonance of the active MRR, then if the power was sufficiently high, parametric gain would eventually result in optical parametric oscillation. Finally Kerr combs with a spacing equal to the MRR FSR were generated $\delta_{OFC}$ (~ 200 GHz). In the wavelength range of the channelizer, the frequency of the $k_{th}$ ($k$=2, 3, 4, …) comb line is

$$f_{OFC}(k) = f_{OFC}(1) + (k-1)\delta_{OFC} \qquad (1)$$

where $f_{OFC}(1)$ is the first comb line frequency on the red side. The combs were then flattened by a Waveshaper and passed through a phase modulator (PM) in order to multicast the broadband RF signal onto all wavelengths. Finally, all of the 200 GHz comb lines, each imprinted with the RF signal spectrum, were spectrally sampled, or sliced, by the second passive high-Q MRR that had an FSR of 49 GHz. Hence, in the first series of experiments we had to use every 4th resonance of the passive MRR to achieve the spectral slicing, which resulted in an effective FSR with a spacing $\delta_{MRR}$ of ~196 GHz. Therefore, the RF spectrum on each of the 200 GHz microcomb lines was progressively sampled with a sequential step between resonances of about 4 GHz. The output channelized RF frequencies are:

$$\begin{aligned}f_{RF}(k) &= f_{MRR}(k) - f_{OFC}(k) \\ &= [f_{MRR}(1) - f_{OFC}(1)] + (k-1)(\delta_{MRR} - \delta_{OFC})\end{aligned} \qquad (2)$$

where $f_{RF}(k)$ is the $k_{th}$ channelized RF frequency, $f_{MRR}(k)$ is the $k_{th}$ centre frequency of the filtering MRR. Here, $[f_{MRR}(1) - f_{OFC}(1)]$ is the relative spacing between the first comb line and adjacent filtering resonance, corresponding to the channelized RF frequency offset, and $(\delta_{OFC} - \delta_{MRR})$ corresponds to the channelized RF frequency step between adjacent wavelengths. While the RF channels could in principle be sampled with an optical demultiplexer such as an arrayed waveguide grating, as long as it had a channel spacing that matched the FSR (or a multiple of) of the 49 GHz MRR, in practice AWG multiplexers and similar devices generally do not have nearly as fine a resolution as the MRRs used here - they tend to have channel widths matched to the ITU grid of 50GHz or 100GHz, versus the MRRs used here that had a 3dB linewidth of about 1GHz. The RF signals were then detected via homodyne detection, where the flattened comb lines were first separated out before phase modulation in order to serve as local oscillators (LO). They were then coupled together with the channelized RF optical sidebands and coherently detected for post processing. In the initial work based on the 200GHz combs [113], however, we only measured the broadband optical spectrum and not the actual RF waveform, to verify the feasibility of our approach, deferring channel demultiplexing and RF detection to later work. Regardless, however, even in that initial work the large number of wavelengths generated by the micro-comb resulted in a large number of channels, with broadband RF channelizing with digital-compatible channel bandwidths.

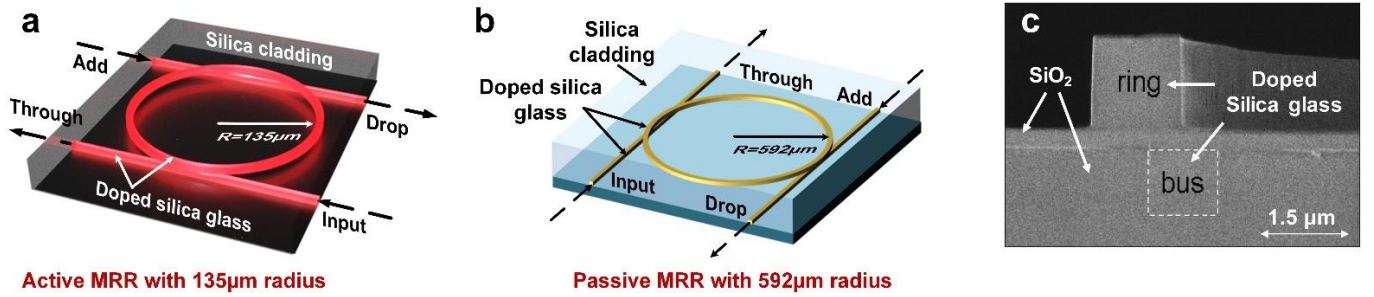

Fig. 2. Schematic illustration of the (a) 200 GHz-FSR MRR and (b) 49 GHz-FSR MRR. (c) SEM image of the cross-section of the 200 GHz MRR before depositing the silica upper cladding.

That first work was followed by an RF channelizer based on a 49GHz microcomb combined with the 49GHz passive MRR filter [122] that resulted in a photonic RF channelizer with significantly improved performance. This was a result of the fact that the two MRRs had quasi-matched FSRs, both near 49GHz. The first MRR generated a soliton crystal microcomb while the second MRR acted as a passive filter. Using the finer spaced (than the original 200GHz) microcomb had two significant benefits. First, the much smaller FSR of the comb source generated as many as 92 wavelength channels across the C band (versus 20 in the initial work). This resulted in a much larger instantaneous operational bandwidth of 8.08 GHz without any thermal tuning – more than 22 times larger than the 200GHz comb device [113]. Secondly, by using an integrated high-Q MRR with a (roughly) matching FSR to slice the RF spectrum via the Vernier effect, an RF channelization step of only 87.5 MHz was achieved. This effectively resulted in continuous RF spectrum channelization since the channel step was comparable to (slightly smaller than) the passive MRR spectral resolution of 121.4 MHz. Finally, another difference with the second device over the first one is that in this device we used parallel phase-modulation to intensity-modulation (PM-IM) conversion across all of the wavelength channels. This directly yielded an RF output in a compact and stable scheme without any need for separate local oscillator paths. Finally, the channelizer's operation frequency range could be dynamically tuned by adjusting the offset frequency between the microcomb and passive MRR. By thermally tuning the passive MRR we succeeded in achieving RF channelization over a very large RF range of 17.55 GHz. In addition to high RF performance, this approach offers a lower complexity, reduced footprint, and even lower cost.

The broadband RF channelizer (Fig. 2 (b)) is very similar to the previous device and also consists of three modules. The first is the microcomb generation and flattening, where an active MRR was pumped by a CW laser to initiate parametric oscillation. With the MRR's high Q factor of $> 10^6$ million, together with the high nonlinear FOM and engineered anomalous dispersion, enough parametric gain was generated to produce Kerr frequency combs. The nature of the oscillation state of the microcomb was determined primarily by the pump to resonance detuning together with the pump power. By sweeping the pump wavelength from blue to red, a range of dynamic nonlinear states could be achieved, including coherent soliton states. The comb lines can also be labelled by Eqs(1,2) except that now the number of lines is much larger. *N* microcomb lines are generated with a spacing of $\delta_{OFC}$, the optical frequency of the $k_{th}$ ($k$=1, 2, 3, …, 92) comb line is given by Eq(1). An optical Waveshaper was again used to flatten the comb line powers.

In the second module, the flattened comb lines were passed through an electrooptical phase modulator, where the input broadband RF signal was multicast onto all wavelengths. Next, as before the replicated RF spectra were sliced by the passive MRR with an FSR of $\delta_{MRR}$, where the slicing resolution is given by the 3dB bandwidth of the passive resonator, about 1GHz in our case. As a result, the RF spectral segments on all wavelength channels were effectively channelized with a progressive RF centre frequency given by Eq(2).

For the 49GHz comb based device, we employed phase modulation and notch filtering (i.e., using the transmission through port of the passive MRR) to convert from phase to intensity modulation. Phase modulation first produced both lower and upper sidebands that had opposite phases. One of the sidebands was then rejected by the notch resonance, leaving the optical carrier to beat with the other unsuppressed sideband on photodetection. Thus, this process resulted in converting the modulation from phase to single-sideband intensity modulation without a local oscillator. Hence, $N$ bandpass filters were achieved, each with a spectral bandwidth resolution $\Delta f$, determined by the passive MRR's resonant linewidth. The center RF frequencies $f_{RF}(k)$ were determined by the spacing between the optical carriers and the passive resonances $[f_{MRR}(k) - f_{OFC}(k)]$, as described in Eq. (2). Therefore, the input RF spectrum was demultiplexed, or channelized, into $N$ segments, each centered at $f_{RF}(k)$ and each having a bandwidth of $\Delta f$.

As mentioned, this approach did not require any other physical local oscillator paths to achieve coherent homodyne detection, and so is much more compact and stable than those involving interfering paths [113]. Finally, the wavelength channels were de-multiplexed and converted back into the electrical domain separately via an array of photodetectors, simultaneously yielding $N$ channelized RF signals, each with a spectral width of $\Delta f$, typically within the operation bandwidth of ADCs. After that, the channelized RF signals were converted into digital signals via an array of ADCs and processed with digital domain tools for further analysis.

# V. 200GHz Microcomb System: Results

Both MRRs (Fig. 2) were fabricated in a high-index doped silica glass platform using CMOS compatible fabrication processes. First, high-index (n = ~1.70 at 1550 nm) Hydex glass films were deposited using standard plasma enhanced chemical vapour deposition (PECVD), then patterned photo-lithographically and etched via reactive ion etching to form waveguides with exceptionally low surface roughness. Finally, silica glass (n = ~1.44 at 1550 nm) or Hydex with the equivalent refractive index, was deposited via PECVD as an upper cladding. The advantages of our platform for nonlinear OPOs include ultra-low linear loss (~0.06 dB·cm$^{-1}$), a moderate nonlinearity parameter (~233 W$^{-1}$·km$^{-1}$), and, in particular, a negligible nonlinear loss up to extremely high intensities (~25 GW·cm$^{-2}$) [38]. The low linear loss resulted in ring resonator $Q$ factors of $> 1\times10^6$. A scanning electron microscope (SEM) image of the cross-section of the 200 GHz MRR before depositing the SiO$_2$ upper cladding is shown in Fig. 2(c). The radii of the active 200GHz MRR (for comb generation) and the passive MRR (for spectral slicing) were ~135 μm and ~ 592 μm, corresponding to FSRs of ~1.6 nm (~200 GHz) and ~0.4 nm (~49 GHz), respectively. After packaging the device with fibre pigtails, the total fibre-fiber through-port insertion loss was ~3.5 dB for the 200 GHz MRR and ~1.5 dB for the 49 GHz MRR.

To generate Kerr combs, the pump power was amplified to ~500 mW and the wavelength swept from blue to red. When the pump wavelength was tuned close to the resonance of the active MRR, primary combs were generated (Fig. 3(a)). Signal and idler lines were generated in the S-band and L-bands respectively, and the spacing between the signal and pump was 19 FSR, or 3.8 THz, determined by the parametric gain curve. When the detuning between the pump wavelength and the adjacent resonance was further changed, the parametric gain lobes broadened and secondary comb lines with a spacing equal to the FSR appeared via cascaded four wave mixing. Finally, flat Kerr combs were generated with the pump wavelength set to 1548.58 nm. As shown in Fig. 3(b–c), the resulting Kerr comb was over 200-nm wide, covering four bands (S, C, L,

U) and was relatively flat over the full C and L bands, thus enabling a record (at the time) large number of wavelengths (20 over the C-band and > 60 over the C/L-bands). The large Kerr comb spacing also yielded an increased Nyquist zone, corresponding to an RF bandwidth of >100 GHz. This bandwidth is extremely challenging for mode-locked lasers and externally-modulated comb sources to achieve.

While the spectral profile of the 200GHz comb was not indicative of operation in the single cavity soliton regime, we did not experience any significant limitations as a result. Theoretical analysis of our observations revealed that the Kerr comb was working under "partially coherent" conditions that featured stable phase and amplitude. In fact, the spatial patterns of what we observed appear to be similar to the so called "soliton molecules" [130,131,171-173], which feature reasonably low intensity noise comparable with that of cavity solitons.

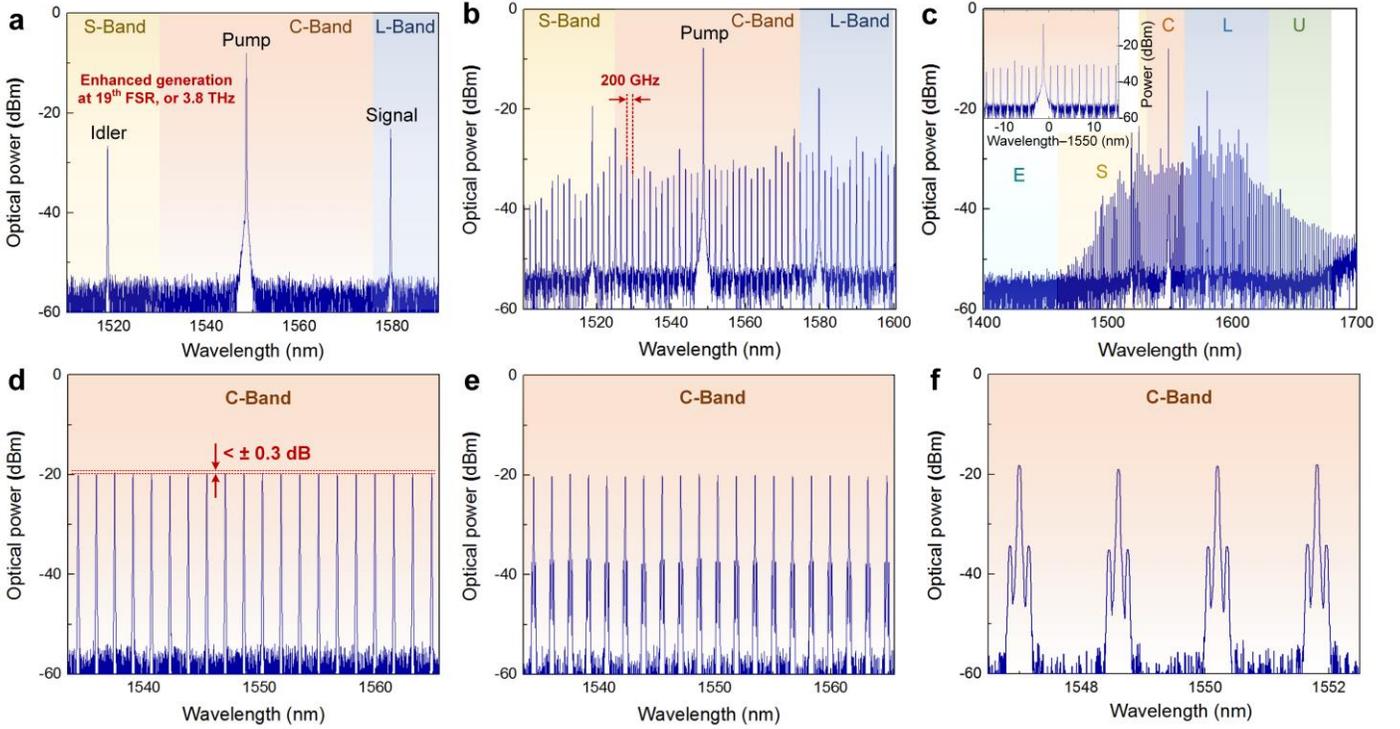

Fig. 3. Optical spectrum for the 200GHz FSR micro-comb device. (a) the primary comb, (b) the secondary comb, (c) the Kerr comb with 300 nm span, (d) the shaped optical comb for the channelizer with less than 0.5 dB unflatness, (e) 20 and (f) selected 4 comb lines modulated by RF signals.

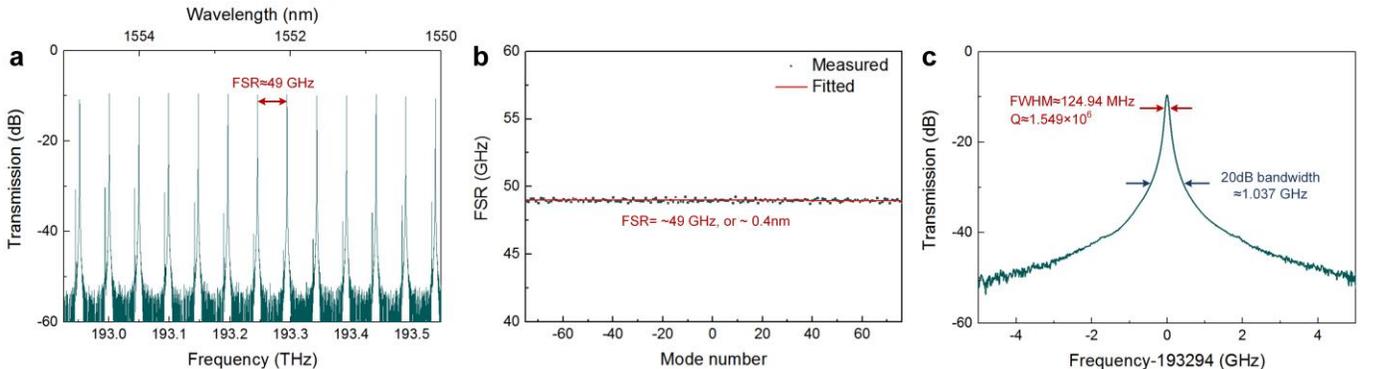

Fig. 4. Drop-port transmission spectrum of the passive on-chip 49 GHz MRR (a) with a span of 5 nm, (b) showing an FSR of 49 GHz, and (c) a resonance at 193.294 THz with full width at half maximum (FWHM) of 124.94 MHz, corresponding to a Q factor of 1.549×10$^6$.

We note that since the comb source could serve as both the optical carrier for RF sideband generation as well as the LO for RF receiving (in the case of the 200GHz comb device), the channelized RF optical sidebands and LOs were inherently coherent, and thus strictly phase-locked comb lines are not necessary — which, in fact, is an advantage when comparing with RF coherent receivers [174]. Similarly, the coherence between adjacent WDM channels is also not required since the different lines do not beat against one another.

The 20 channels over the C-band of the 200GHz Kerr comb were then shaped by a Waveshaper (Finisar 4000S) to get uniform channel weights. A feedback control path was used to accurately read and shape the comb lines' power, first detected by an optical spectrum analyser and then compared with the ideal weights (uniform in our case). This allowed us to generate a feedback error signal to the waveshaper to calibrate the system and achieve accurate comb shaping and flatness of the comb envelope to within ± 0.3 dB (Fig. 3(d)). Note that in practice passive gain flattening filters can be used in place of the Waveshaper to flatten the spectrum. The phase modulator was then used to multicast the RF signal onto each comb line. Figures 3(e–f) show the optical spectra of the flat comb lines modulated by the RF signal. The 20 equalized channels were limited to the C-band purely by the operational range of the waveshaper. In principle the number of wavelengths could easily be increased up to 60 by using a C+L band waveshaper. Next, the multicast RF signal on each wavelength channel was spectrally sliced by the passive 49 GHz MRR. Figures. 4(a–b) show the drop-port transmission of the 49GHz MRR over a 5nm range, with an FSR of ~49 GHz (~0.4nm) over 152 modes (denoting the mode at 193.294 THz as 0). The transmission of a single resonance (Fig. 4(c)) shows a full width at half maximum (FWHM) of 124.94 MHz, corresponding to a Q factor of $1.549 \times 10^6$, and a 20dB bandwidth of 1.04 GHz, which determines the RF resolution. This RF resolution is compatible with state-of-the-art analog-to-digital converters [143], thus leading to digital-compatible broadband RF channelization.

To illustrate the operation principle of the RF channelizer, we show the 200 GHz comb spectrum and transmission of the 49 GHz passive MRR (highlighting every fourth resonance that aligns with the 200 GHz comb) in Fig. 5(a). Fig. 5(a) shows that for the 20 comb channels over the C-band, the relative offset between the 200 GHz comb lines and every fourth 49 GHz MRR resonance yields the RF spectral sliced frequency, which varies from 3.89 to 88.65 GHz. The relative spacing of the 200 GHz comb and 49 GHz MRR were linearly fit (Fig. 5(b)) obtaining a comb spacing of $\delta_{OFC}$ = 200.44 GHz and a fit FSR of the spectrally sliced resonances (i. e., $\delta_{MRR}$ is the frequency spacing between every $4_{th}$ resonance of the 49 GHz MRR) of $\delta_{MRR}$ = 196.01 GHz, thus resulting in a channelized RF frequency step between successive channels ($\delta_{OFC} - \delta_{MRR}$) of 4.43 GHz.

We measured the RF performance of the channelizer up to 20 GHz. This only required 4 of the 20 channels in Figure 5, corresponding to four RF frequencies from 1.7 to 19.0 GHz. We measured the output of the 49 GHz passive MRR with an optical spectrum analyzer (Fig. 6), showing that optical RF sidebands with different RF frequencies were channelized at the different wavelengths. As Fig. 6(a) shows, optical-carrier RF tones at frequencies of 1.7 GHz, 6.3 GHz, 11.2 GHz, and 15.8 GHz were output on 4 wavelength channels, corresponding to an RF frequency step of ~ 4.8 GHz, in agreement with Fig. 5. The power leakage of channel 4 was mainly due to spurious optical carrier tones, which could be reduced by using an optimized modulation format such as carrier-suppressed single-sideband modulation.

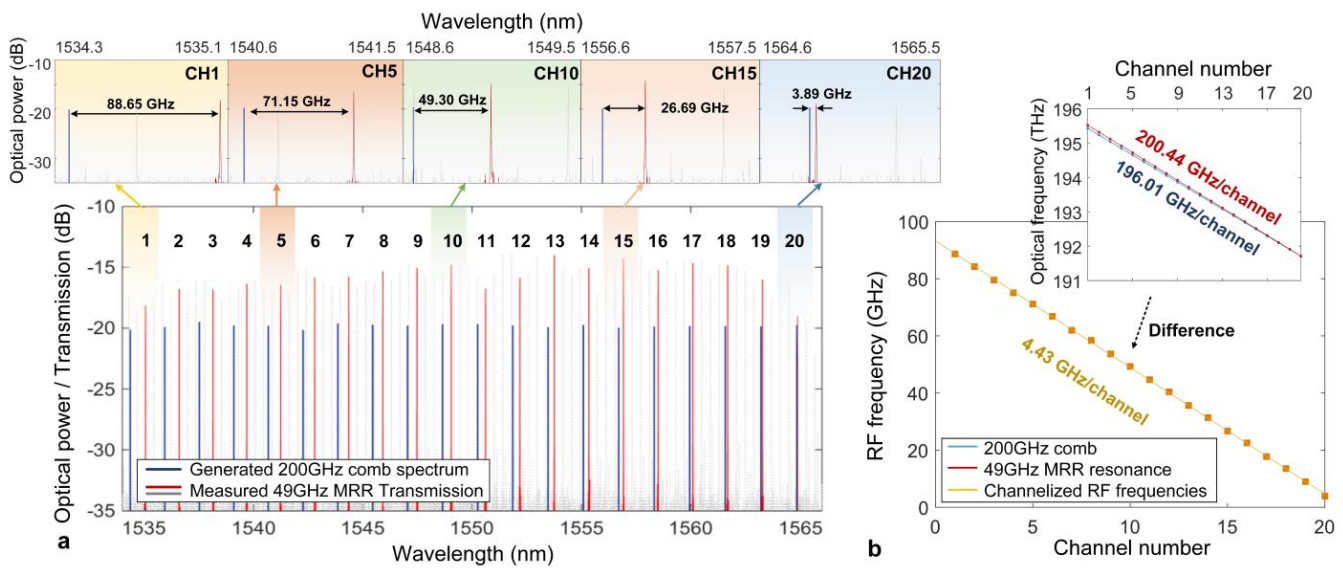

Fig. 5. (a) The measured optical spectrum of the 200GHz micro-comb and transmission of 49GHz MRR. Zoom-in views of the channels with different channelized RF frequencies. (b) Extracted channelized RF frequencies, inset shows corresponding optical frequencies of the comb lines and spectral slicing resonances.

In order to make full use of the RF resolution of our system (~1 GHz) and bridge the gap formed by the ~ 4.8 GHz RF step size, we additionally used temperature tuning [47, 174] of the passive MRR to continuously control the relative spacing between the source and filtering MRRs' resonances ($f_{MRR}(1) - f_{OFC}(1)$), with a millisecond thermal response time. Figures 6(a–d) show the output optical spectra at temperatures from 24.0°C to 25.5°C. As reflected in Fig. 6(e), the channelized RF frequency offset increased at ~1GHz/°C. By combining the fine resolution temperature tuning with the 4.8 GHz RF increment across 60 channels over the C/L-bands, this device could achieve a very wide RF bandwidth of greater than 100 GHz.

Figure 7 shows the channelized RF frequencies of the different optical channels extracted from Fig. 6, over an RF bandwidth from 1.7 to 19.0 GHz, limited by the RF modulator bandwidth. Figure 8 shows the extinction ratio (ER) of the wavelength channels (the optical power ratio of the signal channel to the maximum of other channels) extracted from Figs. 6(a–d), showing up to a 20 dB ER. The non-uniformity in optical power of the different channels mainly arose from the amplifier spectral gain profile, and this can be greatly reduced with gain flattening filters. The extinction ratio was also limited by excess noise of the second EDFA when adapting the optimized RF modulation format to increase the OSNR. In this case, using lower noise EDFAs would enhance the performance. The thermal controllers of the active and passive MRRs had an accuracy of 0.01°C, contributing to an error of ~50 MHz for the channelized RF frequencies, which could be eliminated by using more precise thermal control or a feedback loop to achieve high performance in practical applications. Although employing thermal tuning posed challenges for simultaneous full RF spectrum channelizing, it nonetheless allowed us to demonstrate the capability and the novelty of our method. Thermal tuning is in fact not necessary, provided the RF frequency step between adjacent wavelength channels (i.e., the FSR difference between the active MRR and the passive MRR) is equal to the spectral slicing resolution (i.e, the 20dB bandwidth of the passive MRR), which can be achieved via precise lithographic control of the FSRs of the MRRs. For example, given $\delta_{OFC}$=200 GHz, then designing the passive MRR such that $\delta_{MRR}$=199 GHz, still with a 1 GHz 20dB- bandwidth at the same time. Thus, the full RF spectra could be simultaneously channelized at a resolution of 1GHz without thermal tuning. While the RF operation bandwidth, given by the product of channel number and resolution, would be slightly less than the device reported here, at 60 GHz for 60 wavelengths (channels) across the C/L bands, this could be increased by using a smaller FSR comb (eg., 100 GHz vs. 200 GHz). Finally, we note that our scheme generates RF output based on homodyne detection, whereas down-conversion of the channelized RF signal would require heterodyne detection where a second optical frequency comb with specially designed FSR and offset is required. Recent advances in dual-comb spectroscopy provide new possibilities for that [163, 175-177]. By employing integrated dual micro-combs with specially designed FSRs and offsets for heterodyne detection [178-180], broadband RF signals can be channelized and down-converted into digital bandwidths for direct analog-to-digital conversion and post processing, thus offering a highly competitive approach for integrated photonic RF receivers.

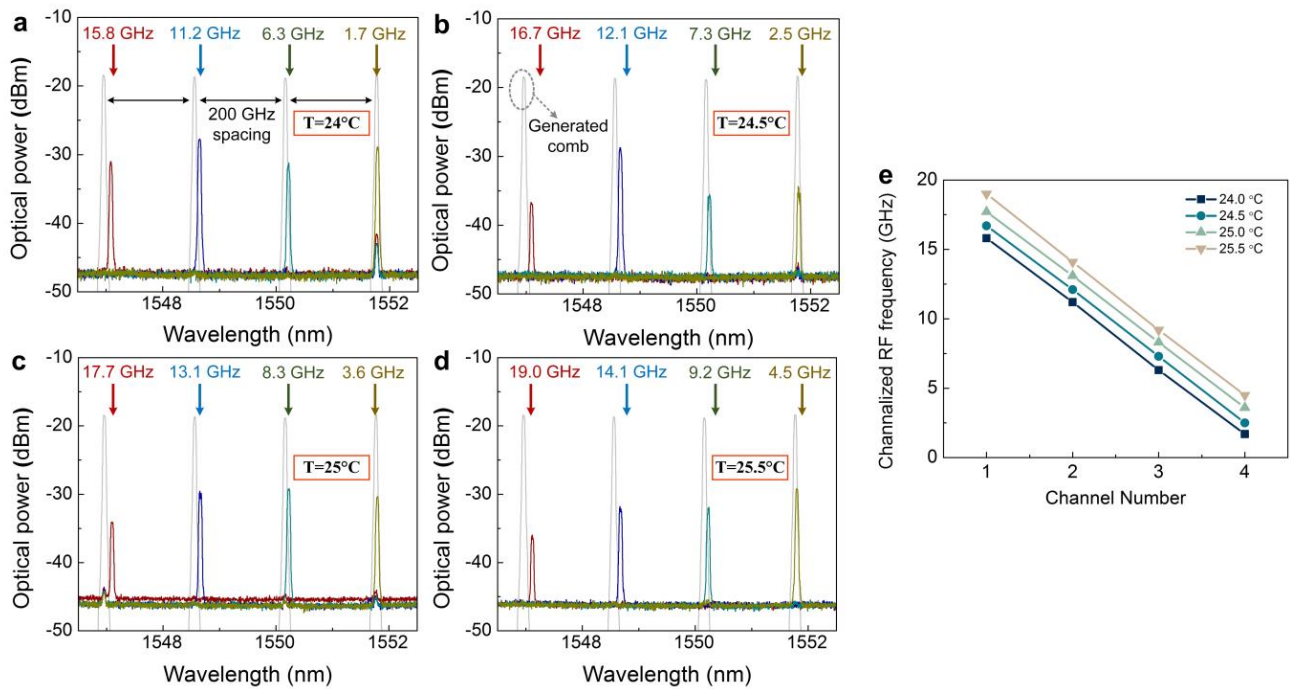

Fig. 6. RF response of 200GHz Channelizer. Measured optical spectrum of 49 GHz MRR's output with different input RF frequencies, with the temperature of the 49 GHz MRR set to (a) 24.0°C, (b) 24.5°C, (c) 25.0°C, and (d) 25.5°C. (e) Channelized RF frequencies at different wavelength channels with different temperatures.

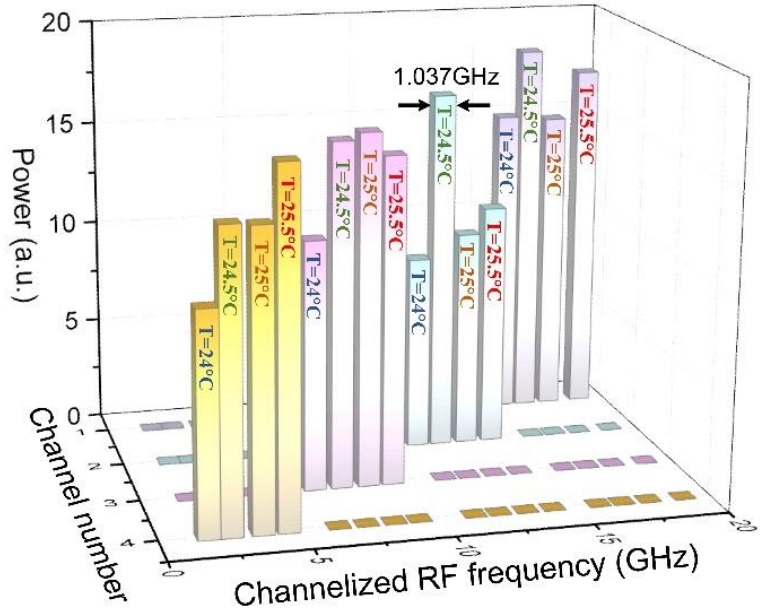

Fig. 7. Channelized RF frequencies at different channels.

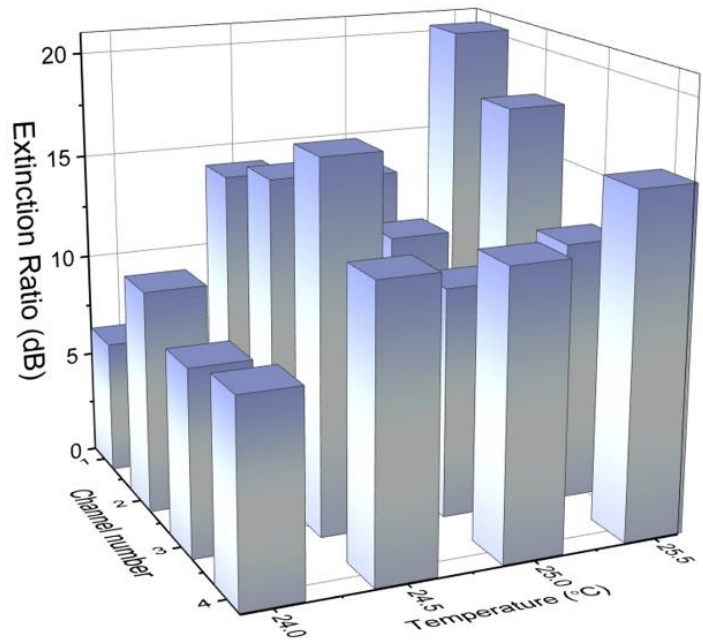

Fig. 8. Extracted extinction ratio of channelized RF signals.

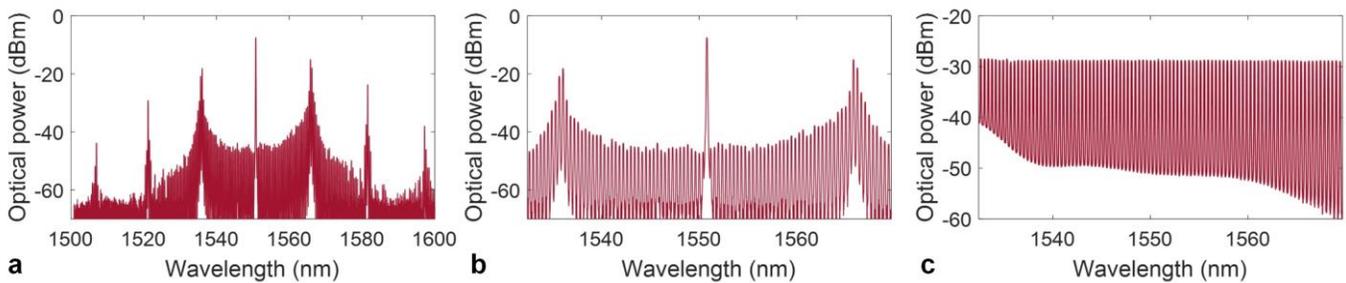

Fig. 9. Optical spectrum of the generated soliton crystal microcomb with (a) 100 nm and (b) 40 nm span. (c) Flattened 92 comb lines.

## VI. 49GHz Microcomb Results

In this section, we summarize the results for the photonic RF channelizer based on two MRRs with quasi-matched FSRs - both near 49GHz. This approach achieves significantly improved performance over the previous device based on a 200GHz spaced microcomb. Here, the first MRR generates a soliton crystal microcomb at 49GHz while the second MRR acts as a passive filter with approximately the same channel spacing. This results in two significant benefits. First, the much smaller overall FSR of the comb source provides up to 92 wavelengths across the C band. This results in a significantly enhanced instantaneous RF bandwidth of 8.08 GHz – more than 22 x that of the previous device [113]. By using integrated high-Q MRRs with approximately matching spacings to both generate the comb and slice the RF spectrum using a Vernier effect, we achieved a RF channelization step between channels of 87.5 MHz. This led to successful continuous channelization of the RF spectrum since the channelization step was smaller than the spectral resolution of 121.4 MHz. We also employed parallel phase-modulation to intensity-modulation (PM-IM) conversion across all wavelength channels — which directly generated an RF output in a stable scheme and compact footprint without needing separate local oscillator paths. Finally, the channelizer's total RF operation frequency range could be dynamically tuned by adjusting the offset frequency between the active microcomb and passive MRR. We thermally tuned the passive MRR to achieve RF channelization over a total bandwidth of 17.55 GHz. In addition to achieving high RF performance, our approach offers lower complexity, a reduced footprint, and potentially even lower cost.

Figure 1b shows the setup of the broadband RF channelizer. As before, the microcomb generation was generated by pumping the first MRR with a CW laser to initiate parametric oscillation. The MRR's high > 1 million Q factor together with the high nonlinear figure of merit, and tailored anomalous waveguide dispersion, all contribute to achieving parametric gain to generate the Kerr frequency combs. The state of operation of the frequency comb was determined by the pump to resonance detuning as well as the pump power. By sweeping the pump wavelength from blue to red, diverse nonlinear dynamic states, including the coherent soliton states, could be triggered. As before, for *N* microcomb lines generated with a spacing of $\delta_{OFC}$, the optical frequency of the $k_{th}$ (*k*=1, 2, 3, …, 92) comb line is given by Equation (1). Again, an optical spectral shaper (Waveshaper) equalized the channel powers. In the second module, the flattened comb lines were modulated with an electrooptical phase modulator to multicast the RF signal onto all wavelengths. Next, the replicated RF signals imprinted on all comb lines were sampled by the passive MRR with an FSR of $\delta_{MRR}$, and with a resolution determined by the 3dB bandwidth of the resonator. As a result, the RF spectral segments on all wavelength channels were channelized with a staggered RF frequency, given by Eq. 2.

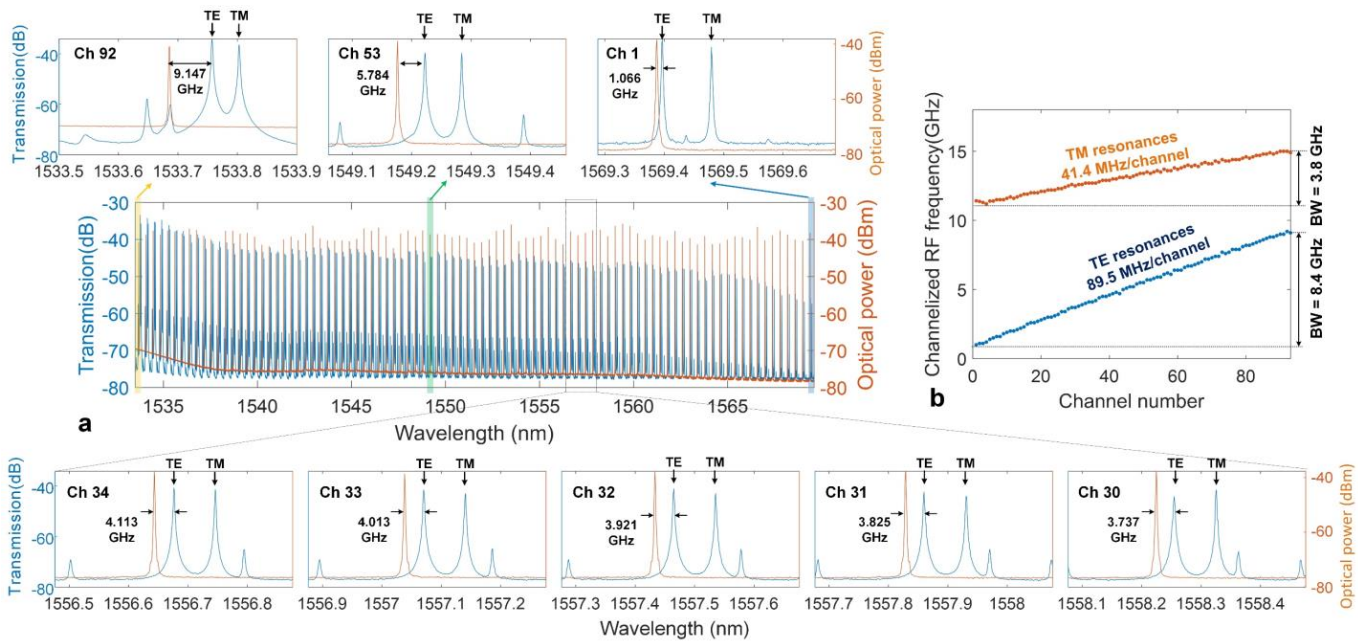

Fig. 10. (a) The measured optical spectrum of the micro-comb and drop-port transmission of passive MRR. (b) Extracted channelized RF frequencies of the 92 channels, calculated from the spacing between the comb lines and the passive resonances. Note that the labelled channelized RF frequencies in (a) is adopted from the accurate RF domain measurement using the Vector Network Analyzer, as shown in the next figure.

For this device we used phase modulation and notch filtering (i.e., the transmission of the passive MRR's through port) to achieve phase to intensity-modulation conversion. The phase modulation first yielded upper and lower sidebands with opposite phases, one of which was then suppressed by the notch resonances, leaving the optical carrier and the other unsuppressed sideband to subsequently beat together upon photodetection. Thus, this process effectively converted the modulation format from phase to intensity modulation (single-sideband). As a result, $N$ bandpass filters were realized, each with a spectral resolution ($\Delta f$) determined by the passive MRR's resonant linewidth. The center RF frequencies $f_{RF}(k)$ are determined by the spacing between the optical carriers and passive resonances [$f_{MRR}(k) - f_{OFC}(k)$], as described in Eq. (2). As a result, the input RF spectrum is channelized into $N$ segments, each centered at $f_{RF}(k)$ and with a bandwidth of $\Delta f$. This approach does not require any physical local oscillator paths to achieve coherent homodyne detection, and so is much more compact and stable than the first approach [113]. Finally, the wavelength channels were de-multiplexed and converted back into the electrical domain separately via an array of photodetectors, simultaneously yielding $N$ channelized RF signals, each with a spectral width of $\Delta f$ within the operation bandwidth of ADCs and then converted into digital signals via an array of ADCs and processed with digital domain into digital signals via an array of ADCs and processed with digital domain tools.

The active and passive MRRs were both fabricated in a CMOS-compatible doped silica glass platform, employing fabrication processes as discussed above. The dispersion of the active 49GHz FSR MRR was tailored to be anomalous in the C band to enable parametric oscillation. Further, a mode-crossing at ~1556 nm was engineered which could initiate the background wave required for soliton crystal generation. During comb generation, the pump power was boosted to ~ 2W while the wavelength was swept manually from blue to red. As the detuning between the pump wavelength and the active MRR's resonance became small enough to ensure sufficient modulation-instability gain in the active MRR, primary combs were initiated. As the detuning was changed further this was followed by soliton crystal microcombs. Distinctive 'palm-like' optical spectra of the soliton crystals were observed, as shown in Fig. 9 (a, b), the curtain-like spectrum a result of the interference between the tightly-packed solitons circulating around the ring [130, 131]. The soliton crystal combs featured a spacing equal to the FSR of the active MRR ~ 48.9 GHz, enabling up to 92 channels in the C-band. Soliton crystals are coherent and low-noise and, due to the ultra-high intracavity power, could easily be generated by manually sweeping the pump wavelength — much easier than the single-soliton states.

Next, the generated soliton crystal microcombs were flattened with a Waveshaper (4000S) to equalize the power of the wavelengths. We used a feedback control path to shape the comb lines' power accurately, which were monitored by an optical spectral analyzer and compared with the desired channel weights to generate an error signal used to program the Waveshaper loss. The flattened comb spectrum is shown in Fig. 9(c). Then the 92 flattened microcomb lines were fed into a phase modulator (iXblue MPZ-LN-20, half-wave voltage = 6V) and served as optical carriers, thus broadcasting the input RF signal onto all wavelengths. The RF replicas were then spectrally sliced by the passive MRR. We employed phase modulation in combination with the notch filters (the through-port transmission of the passive MRR) to map the high-Q resonances of the passive MRR onto the RF domain, after which the input RF spectrum was channelized into N=92 channels each centered at $f_{RF}(k)$ with a bandwidth of $\Delta f$.

To quantify the key experimental parameters mentioned above, we measured the transmission spectrum of the passive MRR (Fig. 10(a)) using a broadband incoherent optical source using the amplified spontaneous emission of an EDFA. Both TE and TM polarized resonances were observed, as shown in the figure. As can be seen, the channelized RF frequency $f_{RF}(k)$, calculated from the spacing between the comb line and the adjacent passive resonance, progressively decreases from the blue to red. The resulting RF channelized frequency (Fig. 10(b)) shows a total operation bandwidth of 8.4 GHz and 3.8 GHz for TE and TM resonances, respectively, with a RF channelization step of 89.5 MHz (TE) and 41.4 MHz (TM) per channel. Considering the channelization resolution, denoted by the linewidth of the passive resonances, is ~120 MHz (measured in the following experiments), we employed the TE passive resonances to obtain a wider bandwidth and lower adjacent-channel interference. By aligning the polarization of the passive MRR's input light with the TE mode, the TM passive resonances were not excited and so did not affect the device performance. The birefringence and dual polarization modes of the MRRs are discussed elsewhere [114, 115].

After the passive MRR, the RF spectrum was channelized into multiple segments carried on different wavelengths which were de-multiplexed into parallel spatial paths and detected separately. We employed an RF Vector Network Analyzer to accurately measure the performance of our channelizer. The measured RF transmission spectra for 92 channels are shown in Fig. 11 (a, b), directly verifying the feasibility of our approach and the realization of 92 parallel RF channels. The imbalanced RF channel power can be equalized by including the RF transmission spectra into the comb feedback shaping loop, where the error signal is generated by the difference between the ideal channel weights with the RF power, instead of the optical power. The signal-to-noise ratio of the channelizer obtained from the RF transmission spectra was 23.7 dB, which can be improved by increasing the extinction of the periodic optical notch filter. For the MRRs, this can be achieved by slightly adjusting the bus waveguide to MRR coupling coefficient.

The centre frequencies of each RF channel (or channelized RF frequencies), were extracted as shown in Fig. 11(c), showing a demonstrated RF channelization step of 87.5 MHz per channel and thus a total instantaneous bandwidth of 8.08 GHz—matching well with our calculations from the MRR's transmission spectrum. The average RF channel resolution, calculated from the 3dB bandwidth of each channel's RF transmission spectrum, was measured to be 121.4 MHz, which was enabled by the passive MRR's high Q factors. Such a high resolution greatly relaxed the requirement of the ADCs bandwidth and indicates that our approach is compatible with a wide range of digital components that typically feature relatively small bandwidths.

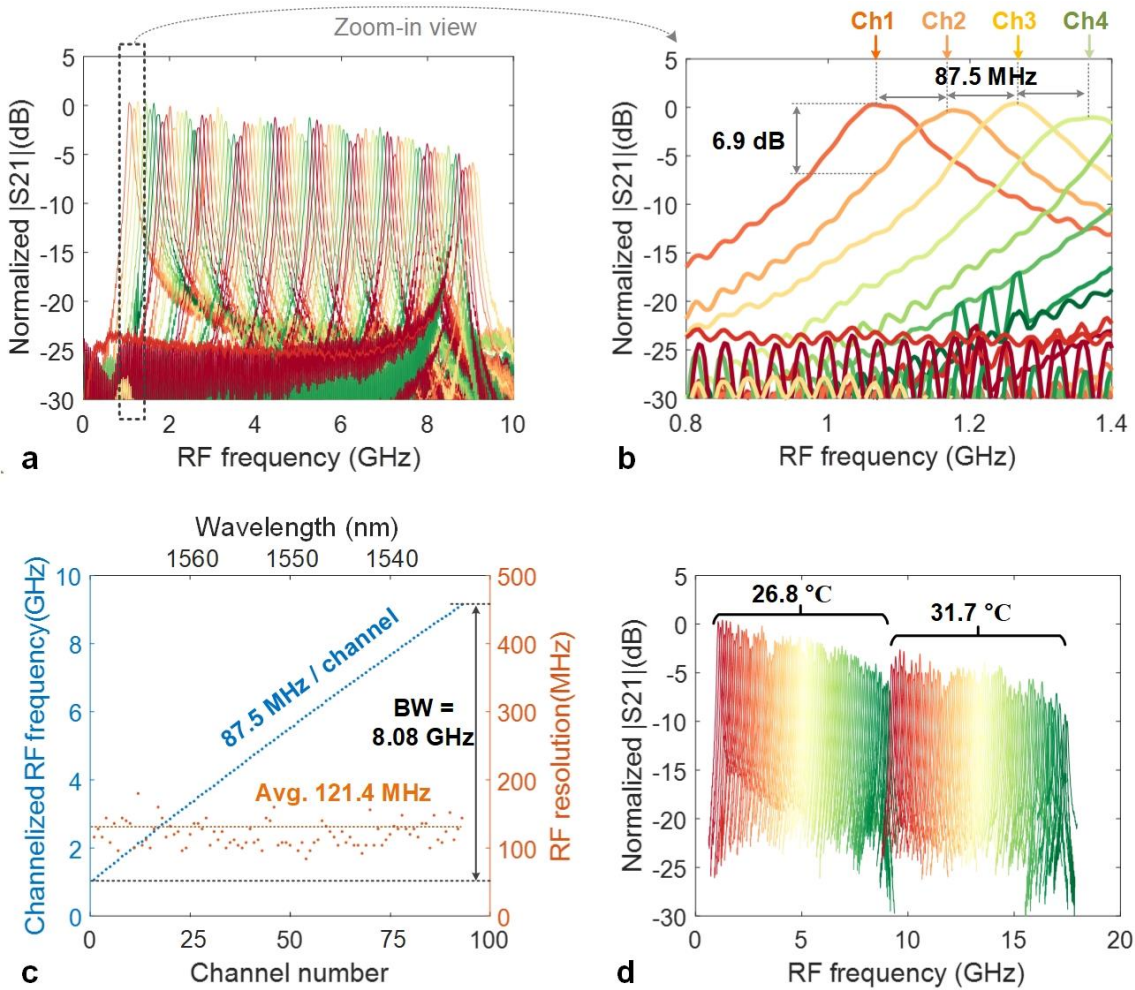

Fig. 11. Measured RF transmission spectra of (a) the 92 channels and (b) a zoom-in view of the first 4 channels. (b) Extracted channelized RF frequency and resolution. (d) Measured RF transmission spectra at different chip temperatures of the passive MRR.

## VII. Discussion

We note that the results for the soliton crystal-based device have a relatively high adjacent-channel crosstalk, which arises predominantly from the mismatch between the RF resolution (121.4 MHz) and the RF channelization step (87.5 MHz) which should ideally be equal. While having a channel step that is smaller than the resolution is not ideal, any resulting unwanted channel crosstalk can be discarded by eliminating redundant channels. Fine tuning the passive MRR's FSR during nanofabrication can increase the RF channelization step to match the RF resolution. This would also lead to a larger instantaneous bandwidth of 121.4MHz×92=11.17 GHz. Further, a high-order optical filter [181] with a flat passband and higher roll-off rate than the Lorentzian lineshape of the single MRR could also reduce adjacent-channel interference. Finally, a slightly higher Q factor ring resonator would also increase the filtering resolution.

Finally, we note that the soliton crystal channelizer's operation band is tunable over a wide range. By thermally tuning the passive MRR, the relative spacing between the source and filtering MRRs' resonances ($f_{MRR}(1) - f_{OFC}(1)$) can be dynamically controlled, where a thermal response time on a millisecond time-scale is expected. As shown in Fig. 11(d), by adjusting the chip temperature of the passive MRR, we shifted the instantaneous operation band of the channelizer from 1.006-9.147 GHz to 9.227-17.49 GHz, continuously covering a total RF bandwidth of 16.48 GHz. We note that, although thermal tuning does change the passive MRR FSR, this effect was very small, in fact ~3955 times (193.4 THz/48.9 GHz) smaller than the change in the relative spacing ($f_{MRR}(1) - f_{OFC}(1)$). The change in the comb line to MRR offset ( ie., $f_{MRR}(1) - f_{OFC}(1)$) that we induced via thermal tuning was 8.221 GHz, while the associated change in the passive MRR FSR (ie., the RF channelization step) was only 2.1 MHz, which is negligible. Therefore, this issue did not impact our device performance.

In addition, although the instantaneous RF operation band could be shifted across a wide range via thermal tuning, the lowest operation RF frequency, which was ~ 600 MHz in our experiments, is limited by a couple of effects. Thermal locking the center resonance wavelength of the filter to the wavelength of the optical carrier [39] places a lower limit on the RF response. Also, the relatively low spectral roll-off of the passive MRR used in our case (with a Lorentzian lineshape) also leads to a deterioration of the phase-to-intensity modulation. This can be addressed by using high-order filters with a steeper roll-off [181-185], while the thermal issues can be minimized by improving the control. The maximum operation frequency of our channelizers was determined by the Nyquist frequency of our microcomb source, which was 48.9 GHz / 2 =24.45 GHz for the soliton crystal devices and 100GHz for the 200GHz microcomb device. Hence, the 200 GHz system had a much higher potential maximum RF operation frequency of 100 GHz—although this comes with a tradeoff with the number of comb lines within the available wavelength range.

## VIII. CONCLUSIONS

We demonstrate broadband RF channelizers based on CMOS-compatible integrated optical frequency comb sources. Broadband 200 GHz-spaced Kerr combs as well as soliton crystal combs with a 49GHz spacing are employed. Both combs provided a record large number of wavelength channels as well as a large RF operation bandwidth. The 200GHz comb-based device has a large potential upper bandwidth of 100GHz, while the soliton crystal device, due to its small (49GHz) spacing that and closely matches the FSRs of the passive MRR filter, up to 92 available wavelength channels were generated, resulting in a broad RF instantaneous bandwidth of 8.08 GHz. A high RF slicing resolution of 121.4 MHz was achieved by the high-Q passive MRR that served as the periodic optical filter for spectra slicing for both devices. For the soliton crystal device, phase to intensity modulation format conversion was employed to ensure stable signal detection without the need for any external local oscillator paths. Dynamic tuning of the RF operation frequency range was achieved through for both devices through thermal control applied to the passive MRR, achieving RF operation up to 17.49 GHz for the soliton crystal device. This device also achieved broadband channelization of RF frequencies from 1.7 GHz to 19 GHz with a high spectral slice resolution of 1 GHz via thermal tuning of a passive on-chip MRR with a Q factor of $1.549 \times 10^6$. These micro-comb based RF channelizers are highly attractive for achieving broadband RF channelization with large channel numbers, high resolution, small footprint, and potentially low cost. They feature massively parallel channels and are highly promising for broadband instantaneous signal detection and processing, representing a solid step towards fully integrated photonic receivers for modern RF systems.